# Coercivity-Aware Machine Learning Discovery of Rare-Earth-Free Soft Magnetic Alloys with First-Principles Magnetic Benchmarks

Avik Mahata[1]*, David Priefer[2]

[1]Department of Mechanical and Electrical Engineering, Merrimack College, North Andover, MA, 01845, USA

[2]Pritzker School of Molecular Engineering, The University of Chicago, Chicago, IL, 60637, USA

**Abstract**

Coercivity-aware machine learning is developed to screen rare-earth-free soft magnetic alloys using Curie temperature and coercivity as coupled design targets. A compiled experimental magnetic-material dataset is evaluated using composition-grouped partitioning to avoid overlap of identical compositions between training and testing, and elemental fractions and composition-weighted elemental descriptors are examined for prediction of the two magnetic properties. Residual-based prediction intervals and distance-to-training filters are then incorporated to screen Fe–Co–Ni–Mn–Al–Si compositions. Composition-grouped evaluation shows lower predictive accuracy than conventional random partitioning, demonstrating the importance of evaluating transfer to previously unseen compositions. Combining weighted elemental descriptors with elemental fractions improves Curie-temperature prediction, whereas the additional descriptors do not improve coercivity prediction, consistent with the strong dependence of coercivity on processing and microstructure. Uncertainty-aware screening substantially narrows the candidate space and identifies Fe–Co-rich compositions for further evaluation. First-principles calculations independently show strong magnetic polarization and magnetovolume effects in representative Fe–Co structures, while finite-temperature atomistic spin simulations place the Fe–Co benchmark systems on high magnetic-ordering-temperature scales. The combined approach provides a computational framework for prioritizing rare-earth-free soft-magnetic chemistries while distinguishing composition-level screening from candidate-specific physical validation.

**Keywords:** soft magnetic alloys; machine learning; composition grouping; prediction intervals; Curie temperature; coercivity

## 1 Introduction

The growing electrification of transportation, energy and industrial systems is increasing demand for magnetic materials that combine efficient flux transport with reliable operation at elevated temperatures. At the same time, concerns over critical-material supply and substitution are shaping materials selection [1–5], further motivating the development of soft magnetic materials for motors, inductors and transformers [6,7]. Soft magnetic performance requires low coercivity together with adequate saturation magnetization, thermal stability and appropriate mechanical properties. The coexistence of mechanical strength and low coercivity has been demonstrated in carefully engineered microstructures [8–11]. Nanocrystalline Fe-based materials, compositionally complex alloys and Fe–Co/Fe–Ni systems provide complementary routes toward these objectives [12–14]. Their properties depend on processing as well as composition, which complicates both experimental optimization and predictive modeling [13,15].

Addressing this coupled dependence on composition, processing and microstructure requires predictive approaches that can efficiently explore large materials design spaces while retaining a connection to the underlying physics. First-principles electronic-structure methods and materials databases provide a foundation for establishing composition–property relationships and identifying promising alloy chemistries [16–19]. For chemically complex alloys, however, predictive modeling must also account for configurational disorder, alloy thermodynamics and competing states [20]. The exact muffin-tin orbital method and coherent potential approximation [21] provide effective-medium descriptions of substitutionally disordered alloys, whereas explicit supercell calculations resolve individual atomic configurations. Magnetic softness introduces an additional level of complexity because coercivity is governed not only by intrinsic magnetic properties but also by magnetization reversal, domain-wall motion, defects and microstructural inhomogeneity [22]. Consequently, quantities such as atomic magnetic moment or Curie temperature alone cannot establish low-coercivity behavior. Explicit supercells represent individual atomic configurations, whereas effective-medium approaches describe an average disordered environment. Both approaches require additional microstructural information to address coercivity in amorphous and nanocrystalline soft magnets [22, 23]

Within this broader modeling framework, different computational methods provide access to complementary aspects of magnetic behavior. Plane-wave density functional theory can quantify magnetic moments, structural energy differences and spin–orbit contributions to magnetic anisotropy [24], providing intrinsic descriptors that can inform the assessment of candidate alloys. Machine learning can complement these calculations by relating such descriptors, together with compositional and microstructural variables, to coercivity when suitable training data are available [25,26]. Atomistic spin models provide another level of description by connecting microscopic magnetic interactions to finite-temperature behavior [27]. These approaches nevertheless describe distinct observables: equilibrium magnetization and magnetization-reversal dynamics are not equivalent [28]. A temperature-dependent magnetization curve therefore provides information on thermal magnetic stability but does not, by itself, constitute a calculation of coercivity. Nachnani et al. combined composition and treatment variables with experimental testing of optimized nanocrystalline alloys [29]. Their work incorporated processing-sensitive descriptors reported $R^2$ values of 0.86 for saturation magnetization and 0.76 for coercivity. A DFT-based alloy dataset achieved $R^2$ values between 0.921 and 0.987 and was coupled with a multi-objective optimization framework [30,31]. For ternary Fe–Si–Al alloys, a coercivity model trained on 122 records using an 85:15 random partition achieved $R^2 = 0.935$, followed by experimental testing of five candidates [32]. For Curie temperature, reported $R^2$ values increased from 0.66 for an uncurated compilation to 0.87 with weighted descriptors on curated data and 0.91 using graph-derived features [33]. Reported accuracies depend on the target property, alloy system, data source and evaluation method. A random partition can place records with identical compositions in both training and testing sets. The resulting score does not measure transfer to compositions absent from training. Meredig et al. illustrated this distinction for logarithmic superconducting transition temperature: $R^2$ decreased from 0.87 under random cross-validation to 0.30 when chemical clusters were held out [34]. Composition-grouped evaluation is therefore useful for assessing models intended to screen new alloy compositions.

In this study, a coercivity-aware machine-learning workflow jointly screens Curie temperature and coercivity using elemental fractions and composition-weighted elemental descriptors. Model performance is evaluated using repeated composition-grouped partitions with nested model selection, followed by residual-based prediction intervals and distance-to-training filters applied to the Fe–Co–Ni–Mn–Al–Si design space. First-principles calculations and finite-temperature magnetic simulations are analyzed independently to provide physical benchmarks for the Fe–Co-rich chemistries emphasized by the screening. Together, these components establish an uncertainty-aware discovery framework that combines statistically robust screening with independent physics-based evaluation of promising soft-magnetic alloy chemistries.

# 2 Methods

## 2.1 Dataset Selection and Provenance

The starting data were obtained from the publicly available Northeast Materials Database (NEMAD), an LLM-assisted compilation of experimentally reported magnetic materials [35]. NEMAD was constructed from scientific articles and historical sources identified using magnetic-material search terms, with structured extraction of chemical composition, magnetic transition temperatures, structural information, and magnetic properties. The original database retained an entry when the reported composition was accompanied by at least one magnetic property or magnetic transition temperature. For the present study, we did not reconstruct the source literature database. Instead, we selected from the publicly released NEMAD data the records relevant to soft-magnet modeling for which both target quantities, Curie temperature and coercivity, were available. The resulting encoded working dataset contains 1,736 records, two target columns, 77 elemental fractions, valence electron concentration, an arithmetic density descriptor, and 27 magnetic-class indicator columns. The subsequent machine-learning analysis therefore represents a task-specific subset of the published NEMAD compilation rather than a new independent experimental database. The encoded file does not retain a unique source identifier, treatment history, microstructure or measurement temperature. It is therefore treated as a compiled property dataset; its entries are not assumed to represent independent experiments or uniformly characterized bulk alloys. Magnetic-class indicators are excluded from the composition-based evaluation and candidate screening because they are unavailable for an unsynthesized composition and may encode information about the target. They are included only in the all-feature random baseline described in Section 2.3.

## 2.2 Composition Descriptors

Three descriptor sets are compared. The elemental-fraction set contains 77 atomic fractions together with the supplied valence electron concentration and arithmetic density descriptors, giving 79 features. For candidate screening, the aggregate descriptors are calculated using Eq. (1), with the same elemental constants and two-decimal rounding used in the encoded training data.

$$\mathrm{VEC} = \sum_{i=1}^{N} x_i v_i, \quad \rho_{\mathrm{desc}} = \sum_{i=1}^{N} x_i \rho_i \tag{1}$$

Here $x_i$ is the atomic fraction, $v_i$ is the adopted valence count and $\rho_i$ is the elemental density. For Fe, Co, Ni, Mn, Al and Si, the valence counts are 8, 9, 10, 7, 3 and 4, and the densities are 7.87, 8.90, 8.90, 7.21, 2.70 and 2.33 g cm$^{-3}$, respectively. The arithmetic density is a numerical descriptor. A physical alloy-density estimate based on additive volumes would also require atomic masses or molar volumes. The second descriptor set follows the composition representation of Ward et al. [36]. Six statistics are calculated for each of 22 elemental properties: atomic number, mass, metallic and covalent radius, Pauling electronegativity, melting and boiling point, density, molar volume, first ionization energy, electron affinity, thermal conductivity, atomization enthalpy, valence electron count and its s, p, d and f contributions, group, period, Mendeleev number and an assigned elemental magnetic moment. The weighted mean, mean absolute deviation and extrema are defined in Eq. (2).

$$p_{\mathrm{mean}} = \sum_{i=1}^{N} x_i p_i, \quad p_{\mathrm{absdev}} = \sum_{i=1}^{N} x_i |p_i - p_{\mathrm{mean}}| \tag{2a}$$

$$p_{\max} = \max p_i, \quad p_{\min} = \min p_i, \quad p_{\mathrm{range}} = p_{\max} - p_{\min} \tag{2b}$$

The mode is the property value of the most abundant element; extrema are taken only over elements with nonzero fractions. Six global descriptors are added: the number of components, largest elemental fraction, ideal mixing entropy divided by the gas constant, effective number of components, atomic-size mismatch and electronegativity mismatch. Together with eight element-family fractions, these give 146 features. The elemental magnetic-moment descriptor is obtained from a fixed lookup table, with zero assigned to unlisted elements; it is therefore used as a compositional feature rather than as a calculated magnetic ground-state property. The tabulated values and assignments are included with the data. Combining both descriptor sets gives 225 features. Coercivity is modeled as the dimensionless logarithm in Eq. (3).

$$y_H = \log_{10}\left(\frac{H_c}{1\ \mathrm{Oe}}\right) \tag{3}$$

### 2.3 Composition-Grouped Evaluation

We compared three partitioning schemes. The all-feature baseline uses an 80:20 random partition of the 1,736-row encoded dataset with the supplied magnetic-class indicators and fixed model settings. The second scheme uses an 80:20 random partition of the cleaned data with composition descriptors only. The third holds out 20% of composition groups, ensuring that an encoded composition cannot occur in both training and testing. The grouped scheme is repeated twenty times with independent partitions. Within each repeat, a further grouped split of the training partition produces an internal validation set that selects both the descriptor set and the learner from nine prespecified combinations: three descriptor sets crossed with random-forest, Extra-Trees and gradient-boosted ensemble models. The selected combination is refitted on the full outer training partition and applied once to the held-out compositions. The outer partition therefore never informs the choice of descriptors, learner or interval width. Performance is summarized by $R^2$, mean absolute error and root mean squared error, together with a training-mean baseline. For observed values $y_i$, predictions $\hat{y}_i$ and test-set mean $\bar{y}$, the metrics are

$$R^2 = 1 - \frac{\sum_{i=1}^{n}(y_i - \hat{y}_i)^2}{\sum_{i=1}^{n}(y_i - \bar{y})^2} \tag{4a}$$

$$\text{MAE} = \frac{1}{n}\sum_{i=1}^{n}|y_i - \hat{y}_i| \tag{4b}$$

$$\text{RMSE} = \sqrt{\frac{1}{n}\sum_{i=1}^{n}(y_i - \hat{y}_i)^2} \tag{4c}$$

Scores are reported as the mean and standard deviation over twenty grouped partitions. Pooled metrics are also calculated from the concatenated test predictions; repeated appearances of a record are prediction instances, not independent observations. A representative reference partition is evaluated under the same nested protocol. The final screening models use the same fixed model family and settings selected by the repeated grouped evaluation; implementation and software-version details are provided with the released code.

### 2.4 Residual-Based Prediction Intervals

Absolute residuals on the internal grouped validation set determine the interval half-width. The empirical construction is motivated by residual-based conformal prediction [37], but the present selection and refitting procedure differs from standard split conformal inference. For m validation records and nominal level $1 - \alpha$, the residuals and quantile level are defined in Eq. (5a), and the interval in Eq. (5b).

$$r_j = |y_j - \hat{y}_j|, \quad \beta = \min\left(1, \frac{\lceil (m + 1)(1 - \alpha) \rceil}{m}\right) \tag{5a}$$

$$q_{1-\alpha} = Q_\beta(r_1, \dots, r_m), \quad C(x) = [\hat{y}(x) - q_{1-\alpha}, \hat{y}(x) + q_{1-\alpha}] \tag{5b}$$

Q denotes the empirical quantile with linear interpolation between ordered residuals, as implemented by NumPy. The validation set is used for model selection, and the selected model is subsequently refitted on the full outer training partition. Consequently, these intervals are evaluated empirically and do not have a distribution-free split-conformal coverage guarantee.

Coverage is measured on untouched outer-test compositions. A reference-partition coverage curve is also compared with intervals defined by the random-forest prediction plus or minus twice the tree-to-tree standard deviation. Repeated records within composition groups and distribution shifts during screening further limit coverage interpretation.

## 2.5 Candidate Screening

A total of 20,000 Fe–Co–Ni–Mn–Al–Si compositions are sampled from a Dirichlet distribution weighted toward Fe–Co-rich compositions while retaining variation in Ni, Mn, Al and Si. Fractions are rounded to three decimal places and renormalized. The screening models use the descriptor set and learner most frequently selected in the repeated grouped evaluations. Fixed Extra-Trees models are fitted on 80% of composition groups, and residuals from the remaining 20% determine the interval widths. These calibration groups are excluded from the final fitting step but were available during the earlier model-selection procedure; their inclusion is therefore not interpreted as independent test coverage. Exact sampling and implementation details are available with the released code. Each candidate receives predictions and residual-based intervals for both targets. Its distance to the nearest fitting composition is calculated as the sum of absolute differences over the 77 elemental fractions. The median and 90th percentile of the corresponding calibration-set distances define close, intermediate and distant categories. These labels describe relative distance from the fitting data rather than demonstrated interpolation or extrapolation. Candidates are selected when their lower Curie-temperature bound exceeds the temperature threshold, their upper coercivity bound is below the coercivity threshold, and their distance is no greater than the calibration-set 90th percentile. Point criteria of $T_C > 1{,}000$ K and $H_c < 100$ Oe are examined, together with a grid of threshold pairs. Because the two targets are treated with separate marginal intervals, two nominal 90% intervals do not imply 90% joint coverage. Among candidates satisfying the specified screening criteria, candidates were ranked in descending order of predicted Curie temperature; coercivity and composition distance were used as acceptance filters rather than ranking variables.

## 2.6 First-Principles Calculations and Energy Normalization

The first-principles calculations use the projector augmented-wave (PAW) method within a plane-wave density-functional framework [38–40], with the PBE exchange-correlation functional and PAW datasets for Fe, Co, Ni, Mn, Al, and Si as applicable. Total energies, cell volumes, and total and site-resolved magnetic moments are extracted from the converged calculations. Electronic convergence is verified before quantitative analysis, and incomplete or duplicate calculations are excluded. The static volume scans comprise seven lattice parameters each for bcc Fe, B2 FeCo, $L1_2$ $Fe_3Co$ and $L1_2$ $FeCo_3$. The bcc/B2 calculations use k-point meshes between $15 \times 15 \times 15$ and $16 \times 16 \times 16$, while the $L1_2$ calculations use meshes between $12 \times 12 \times 12$ and $13 \times 13 \times 13$. Atomic positions and cell vectors are fixed for each sampled lattice parameter. The smallest-volume bcc Fe calculation uses an antiparallel initial spin configuration together with different numerical settings and is therefore excluded from the ferromagnetic energy curve. Because the volume scans are discrete and exhibit changes in magnetic moment, we report the lowest sampled energies rather than fitted equilibrium lattice parameters or elastic properties. The three candidate BCC/FCC pairs contain 54 and 108 atoms and use 5 × 5 × 5 and 4 × 4 × 4 k-point meshes, respectively. Energies are normalized by the atom count of each structure and expressed as $E\sigma\rightarrow0$ per atom; total magnetic moments are normalized similarly. For two phases of identical composition, the energy difference per atom is,

$$\Delta E_{FCC-BCC} = \frac{E_{FCC}}{N_{FCC}} - \frac{E_{BCC}}{N_{BCC}} \quad (6)$$

Such a comparison is physically meaningful only when the chemical composition, numerical treatment, and relevant structural and magnetic states are matched. The available BCC/FCC pairs differ slightly in composition and represent fully relaxed configurations. Their normalized energy differences are therefore reported only as descriptive comparisons and are not interpreted as formation energies or thermodynamic phase preferences. No convex-hull, phonon, or mechanical-stability conclusions are drawn.

## 2.7 Finite-temperature atomistic spin simulations

Finite-temperature atomistic spin simulations were performed for bcc Fe, B2 FeCo, $L1_2$ $Fe_3Co$, $L1_2$ $FeCo_3$, fcc Co, fcc Ni and fcc Mn. These systems were selected as representative magnetic benchmarks for the Fe–Co–Ni–Mn–Al–Si composition space rather than as exact replicas of the machine-learning candidates. The Fe–Co structures probe the dominant Fe–Co-rich chemistry identified by the screening, while Co, Ni and Mn provide additional elemental magnetic references; Mn is treated separately because its net magnetization reflects antiferromagnetic rather than ferromagnetic ordering. Atomistic spin modeling provides a finite-temperature framework for tracking the loss of magnetic order , with Monte Carlo sampling based on the Metropolis approach [27,41,42]. Periodic boundary conditions were used with finite simulation cells containing 4,394 atoms for the BCC systems and 8,788 atoms for the FCC/$L1_2$ systems. Temperature-dependent normalized magnetization was obtained by Monte Carlo sampling after equilibration, with detailed simulation parameters provided in the released computational workflow. The Fe, FeCo, Co and Ni calculations use 5 K temperature spacing, while fcc Mn uses 2 K spacing. The simulated temperature ranges extend beyond the loss of ferromagnetic order for the Fe-, Co- and Ni-based benchmarks. A consistent transition-scale indicator, $T_{0.1}$, is defined as the first sampled temperature at which the normalized magnetization decreases to 0.1 or below. This criterion is used only to compare the finite-temperature magnetic response of benchmark systems; it is not treated as a thermodynamic Curie-temperature estimator because susceptibility peaks, Binder cumulants, finite-size scaling and independent-run uncertainties are not available.

# 3 Results and discussion

## 3.1 Dataset Characteristics and Composition Overlap

Figure 1 summarizes the distributions and internal variability of the two target properties. The cleaned dataset spans 1.85–1,394.15 K in Curie temperature and 0.01–90,000 Oe in coercivity (Figure 1a,b). Because the compilation contains ferrimagnetic, antiferromagnetic, and other magnetic classes in addition to metallic ferromagnets, its 77-element chemical coverage should not be interpreted as a homogeneous alloy training domain. As shown in Figure 1c, valence electron concentration is only weakly correlated with Curie temperature (Pearson correlation = 0.258), indicating that this single descriptor does not provide a mechanistic explanation for the screening trends. Repeated encoded compositions also reveal substantial property variation that cannot be resolved from composition alone. Among the 164 composition groups containing more than one retained record, the median Curie-temperature range is 42 K and the median logarithmic coercivity range is 0.623, corresponding to a factor of approximately 4.2 in coercivity. Figure 1d shows that most repeated groups have relatively modest coercivity variation, while a smaller number span several logarithmic units. Such within-composition differences may arise from processing, sample form, measurement conditions, parsing ambiguities, or transcription errors. Consequently, identical composition descriptors can be associated with different target values, emphasizing an intrinsic limitation of composition-only prediction for this compiled dataset. This composition repetition also affects model evaluation. The all-feature random baseline gives $R^2$ = 0.739 and MAE = 83.15 K for Curie temperature, and $R^2$ = 0.671 and MAE = 0.498 logarithmic units for coercivity. Of the 348 test records, 172 have an encoded composition also present in the training set. After cleaning and exclusion of the magnetic-class indicators, 150 of 343 records in the composition-only random test set still share compositions with training. The higher random-partition scores therefore include substantial composition overlap and do not directly measure transfer to previously unseen compositions.

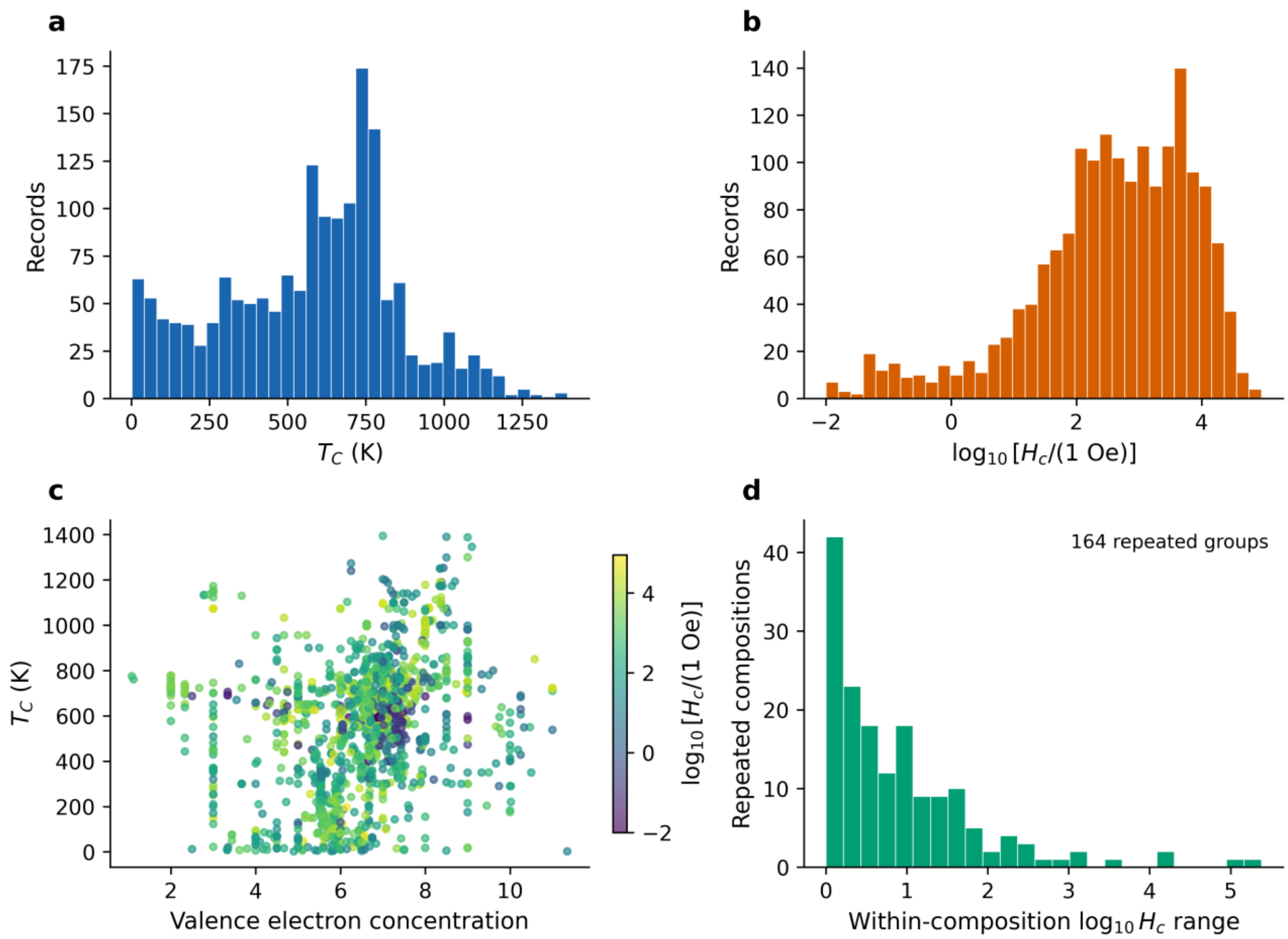

**Figure 1.** (a) Curie-temperature distribution. (b) Logarithmic coercivity distribution. (c) Valence electron concentration versus Curie temperature, colored by logarithmic coercivity. (d) Coercivity variation within the 164 repeated encoded-composition groups. The plotted records include several magnetic-material classes.

Composition-grouped evaluation removes this exact-composition overlap (Figure 2a). Across twenty grouped partitions, $R^2$ ranges from 0.466 to 0.738 for Curie temperature and from 0.425 to 0.693 for logarithmic coercivity (Figure 2b), with corresponding MAEs of 114.0 ± 13.2 K and 0.591 ± 0.051 logarithmic units. The Curie-temperature reference partition gives $R^2 = 0.327$, a test-target standard deviation of 248.9 K, and an RMSE of 203.9 K. Its target spread is the smallest among the examined partitions, while its RMSE exceeds that of 18 of the 20 repeated partitions. Both the narrower target distribution and the larger prediction error therefore contribute to the lower $R^2$ value (Figure 2c). Figure 2d confirms that exact-composition overlap is eliminated in the grouped test sets. The variation between partitions indicates that a single grouped result is insufficient to characterize performance on this dataset. Mean scores and standard deviations over repeated partitions are therefore used for the main comparison. These standard deviations describe partition sensitivity; they are not confidence intervals for performance on a new material population.

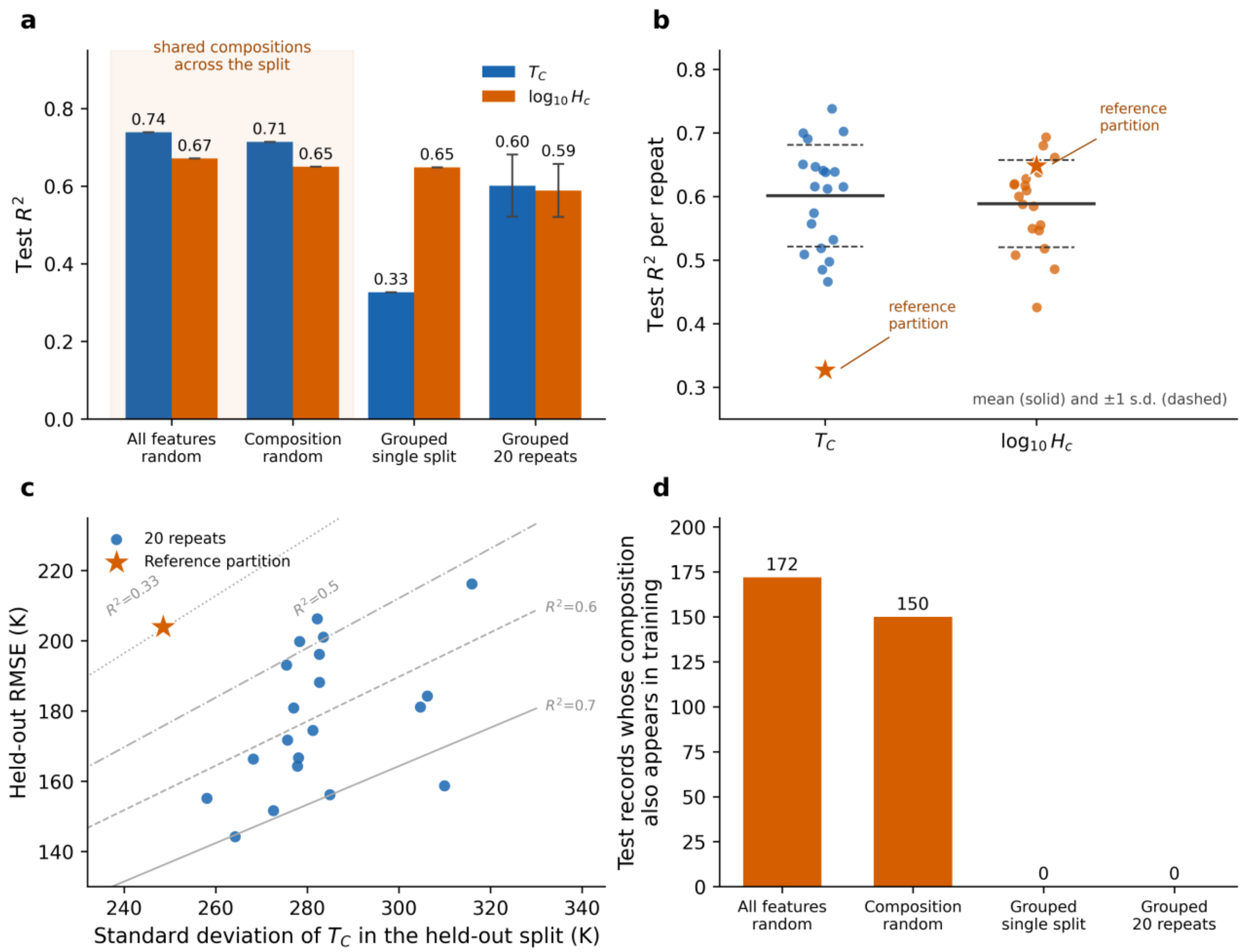

**Figure 2.** (a) Test R² by protocol; shaded protocols share encoded compositions between training and testing. Error bars show the standard deviation over twenty grouped partitions. (b) Individual grouped-partition scores and the representative reference partition. (c) Curie-temperature RMSE against test-target standard deviation with contours of constant R²; the standard deviation uses the population convention for the contour relation. (d) Number of test records sharing an encoded composition with training.

### 3.2 Prediction Accuracy and Descriptor Comparison

The nested evaluation gives $R^2$ = 0.601 ± 0.080 for Curie temperature and 0.589 ± 0.069 for logarithmic coercivity (Table 1). Pooling the test predictions gives $R^2$ = 0.602 and 0.593, respectively. The 6,936 prediction instances represent 1,686 distinct records from 1,043 composition groups, with records appearing in more than one test partition. Mean absolute errors are 114.0 K and 0.591 logarithmic units, compared with 230.8 K and 1.038 for the corresponding training-mean baselines. Figures 3a and 3b show the pooled predictions for Curie temperature and logarithmic coercivity, respectively.

**Table 1.** Test performance by partitioning protocol. The first two protocols include composition overlap. The reference partition and twenty repeated partitions use nested selection of descriptor set and learner. The repeated values are mean ± standard deviation. Pooled metrics use prediction instances rather than independent records. Coercivity errors refer to $\log_{10}[H_c/(1\ \text{Oe})]$.

| Protocol | Target | n test | Test records with shared composition | R² | MAE | RMSE |
|---|---|---|---|---|---|---|
| All-feature random | $T_C$ (K) | 348 | 172 | 0.739 | 83.152 | 141.568 |
| All-feature random | $\log_{10} H_c$ | 348 | 172 | 0.671 | 0.498 | 0.716 |
| Composition random | $T_C$ (K) | 343 | 150 | 0.714 | 93.195 | 146.926 |
| Composition random | $\log_{10} H_c$ | 343 | 150 | 0.650 | 0.513 | 0.731 |
| Grouped reference | $T_C$ (K) | 358 | 0 | 0.327 | 120.063 | 203.923 |
| Grouped reference | $\log_{10}$ Hc | 358 | 0 | 0.648 | 0.597 | 0.849 |
| Grouped, 20 repeats | $T_C$ (K) | 347 (mean) | 0 | 0.601 ± 0.080 | 113.953 ± 13.213 | 177.804 |
| Grouped, 20 repeats | $\log_{10} H_c$ | 347 (mean) | 0 | 0.589 ± 0.069 | 0.591 ± 0.051 | 0.835 |
| Grouped, pooled predictions | $T_C$ (K) | 6936 | 0 | 0.602 | 114.138 | — |
| Grouped, pooled predictions | $\log_{10} H_c$ | 6936 | 0 | 0.593 | 0.590 | — |

A separate comparison of descriptor sets over ten identical grouped partitions is reported in Table 2 and Figure 3c. For Curie temperature, elemental fractions give $R^2$ = 0.600 ± 0.062, weighted elemental statistics give 0.608 ± 0.067, and the combined descriptor set gives 0.630 ± 0.062. The combined set improves the score in seven of the ten paired comparisons and is selected 13 times in the twenty-partition nested evaluation. Elemental statistics therefore provide additional information for Curie-temperature prediction within the evaluated models. For logarithmic coercivity, elemental fractions give $R^2$ = 0.587 ± 0.082, compared with 0.557 ± 0.083 for weighted elemental statistics and 0.565 ± 0.081 for the combined set. Elemental fractions are selected in 12 of the 20 nested evaluations. The absence of consistent improvement from the additional descriptors is consistent with the strong dependence of coercivity on grain size, defects, anisotropy and processing, none of which is explicitly represented by the composition descriptors.

**Table 2.** Descriptor comparison over ten identical grouped partitions, reported as mean ± standard deviation. The selection counts are from the separate twenty-partition nested evaluation in which both descriptor set and learner are selected.

| Descriptor set (features) | Target | Held-out R² | MAE | Times selected |
|---|---|---|---|---|
| Elemental fractions (79) | $T_C$ (K) | 0.600 ± 0.062 | 113.604 | 5 / 20 |
| Elemental fractions (79) | $\log_{10} H_c$ | 0.587 ± 0.082 | 0.581 | 12 / 20 |
| Weighted elemental statistics (146) | $T_C$ (K) | 0.608 ± 0.067 | 111.481 | 2 / 20 |
| Weighted elemental statistics (146) | $\log_{10} H_c$ | 0.557 ± 0.083 | 0.606 | 4 / 20 |
| Both (225) | $T_C$ (K) | 0.630 ± 0.062 | 108.185 | 13 / 20 |
| Both (225) | $\log_{10} H_c$ | 0.565 ± 0.081 | 0.601 | 4 / 20 |

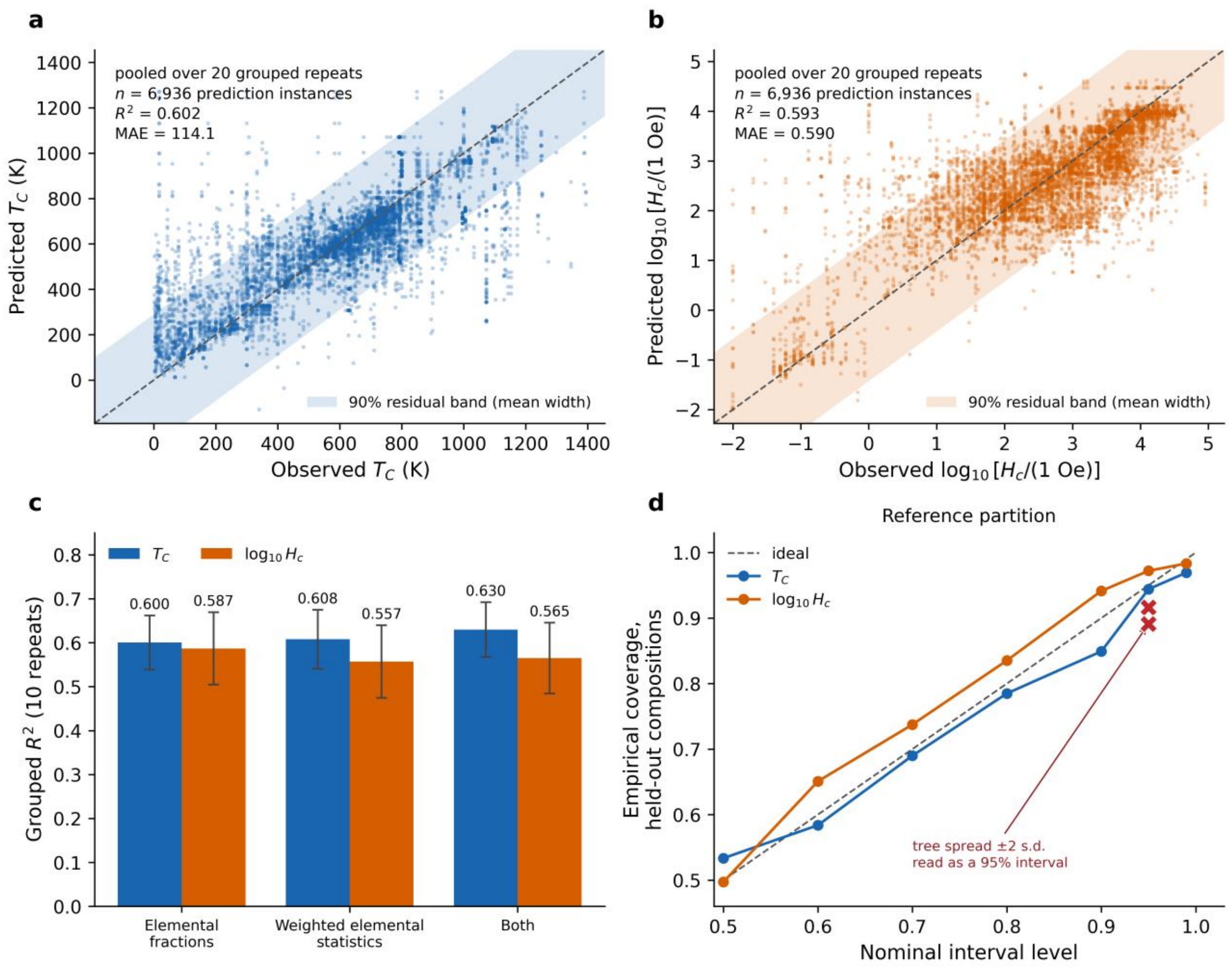

**Figure 3.** Test accuracy, descriptor comparison and empirical interval coverage. (a,b) Pooled test predictions over twenty grouped partitions; shading indicates the mean nominal 90% residual half-width. (c) R² by descriptor set over ten grouped partitions. (d) Coverage on the reference partition for fixed Extra Trees models, compared with separate random-forest intervals using twice the tree standard deviation. Panel (d) is not the mean coverage of the twenty nested evaluations.

The descriptor-attribution results in Figure 4 provide further insight into the information learned by the models. For Curie temperature, the leading SHAP attributions include the Fe, Co and combined Fe+Co+Ni fractions, together with descriptors associated with 3d and valence-electron character (Figure 4a). Permutation importance is dominated by the Fe fraction (Figure 4c), indicating that Fe content provides a particularly strong source of predictive information within this dataset. For logarithmic coercivity, the leading attributions include Sm, Nd, P and B fractions (Figure 4b), while the permutation analysis similarly assigns substantial importance to these compositional variables (Figure 4d). These patterns suggest that the coercivity model is partly distinguishing chemical families, including rare-earth-containing intermetallic magnets and metalloid-containing amorphous or nanocrystalline materials, rather than learning a direct magnetization-reversal mechanism. The resulting chemical-family recognition can aid candidate ranking but cannot distinguish different processing routes or microstructures associated with the same nominal composition.

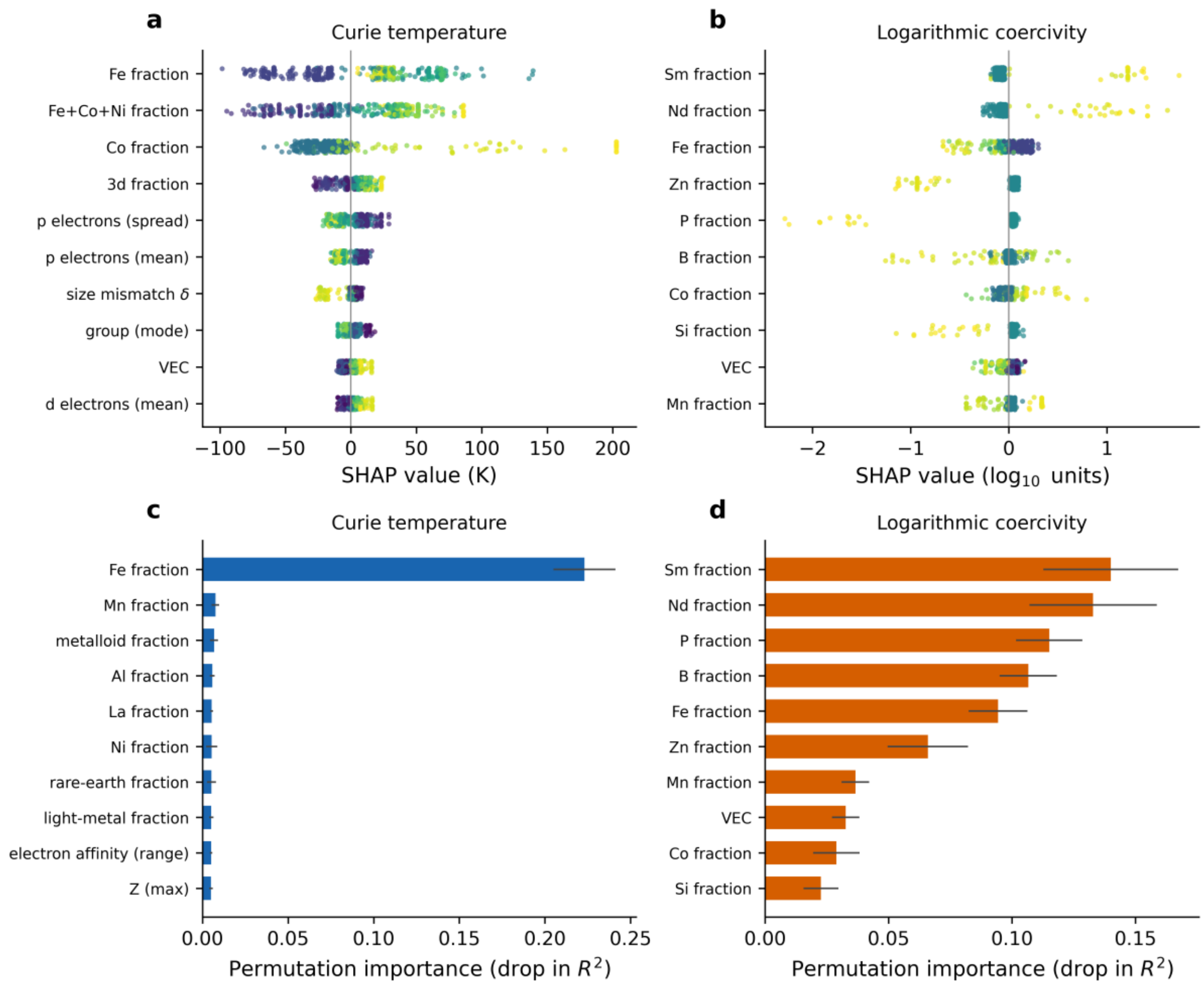

**Figure 4.** Descriptor attributions computed on held-out compositions. (a,b) SHAP values for the ten most influential descriptors for Curie temperature and logarithmic coercivity, colored by the rank of the descriptor value. (c,d) Permutation importance on the same held-out partition, as the drop in $R^2$ with error bars over fifteen permutations. Attributions describe model behaviour, not a physical mechanism.

### 3.3 Empirical Interval Coverage

At the nominal 90% level, the validation-residual intervals achieve mean outer-test coverage of 0.910 ± 0.028 for Curie temperature and 0.908 ± 0.039 for logarithmic coercivity. These values indicate that the empirical intervals provide approximately the intended coverage across the repeated composition-grouped evaluations. For the reference partition shown in Figure 3d, the corresponding coverages are 0.849 and 0.941, respectively, illustrating the variation in coverage between individual partitions. The final screening models yield interval half-widths of 276.9 K for Curie temperature and 1.209 logarithmic units for coercivity, the latter corresponding to a factor of approximately 16.2 in $H_C$. These relatively broad intervals substantially constrain candidate selection and emphasize the uncertainty associated with predictions for unseen compositions. Because the intervals are calibrated using validation residuals, their calibration-set coverage is not interpreted as independent evidence of performance on new candidate materials. For comparison, intervals based on twice the tree-to-tree standard deviation of the reference random-forest models cover 91.6% of the Curie-temperature observations and 89.1% of the coercivity observations (Figure 3d). Tree-to-tree spread therefore does not, by itself, establish nominal 95% coverage. More generally, the residual-based intervals provide an empirical measure of predictive uncertainty, but their coverage may vary with model selection, refitting, and shifts between the training and screened composition spaces.

### 3.4 Screening of Fe Co Ni Mn Al Si compositions

Of the 20,000 screened compositions, only 11 satisfy the point-prediction criteria of $T_C > 1,000$ K and $H_C < 100$ Oe (Figure 5a). None, however, satisfies both interval criteria at these thresholds. The highest-ranked point-prediction candidate has a predicted Curie temperature of 1,070 K with a nominal 90% residual interval of 793-1,347 K. The intervals of all ten highest-ranked candidates cross the 1,000 K threshold (Figure 5b), demonstrating that the point predictions alone do not resolve candidate acceptance at this target.

The screened compositions also differ in their proximity to the fitting data. The median $L_1$ distance to the nearest fitting composition is 0.063 for the calibration set and 0.358 for the generated candidates (Figure 5c). Thus, many generated compositions lie farther from the fitting data than typical calibration compositions. The distance filter excludes candidates beyond the 90th percentile of the calibration-set distance distribution, but it does not by itself establish interpolation, prediction accuracy, or physical feasibility.

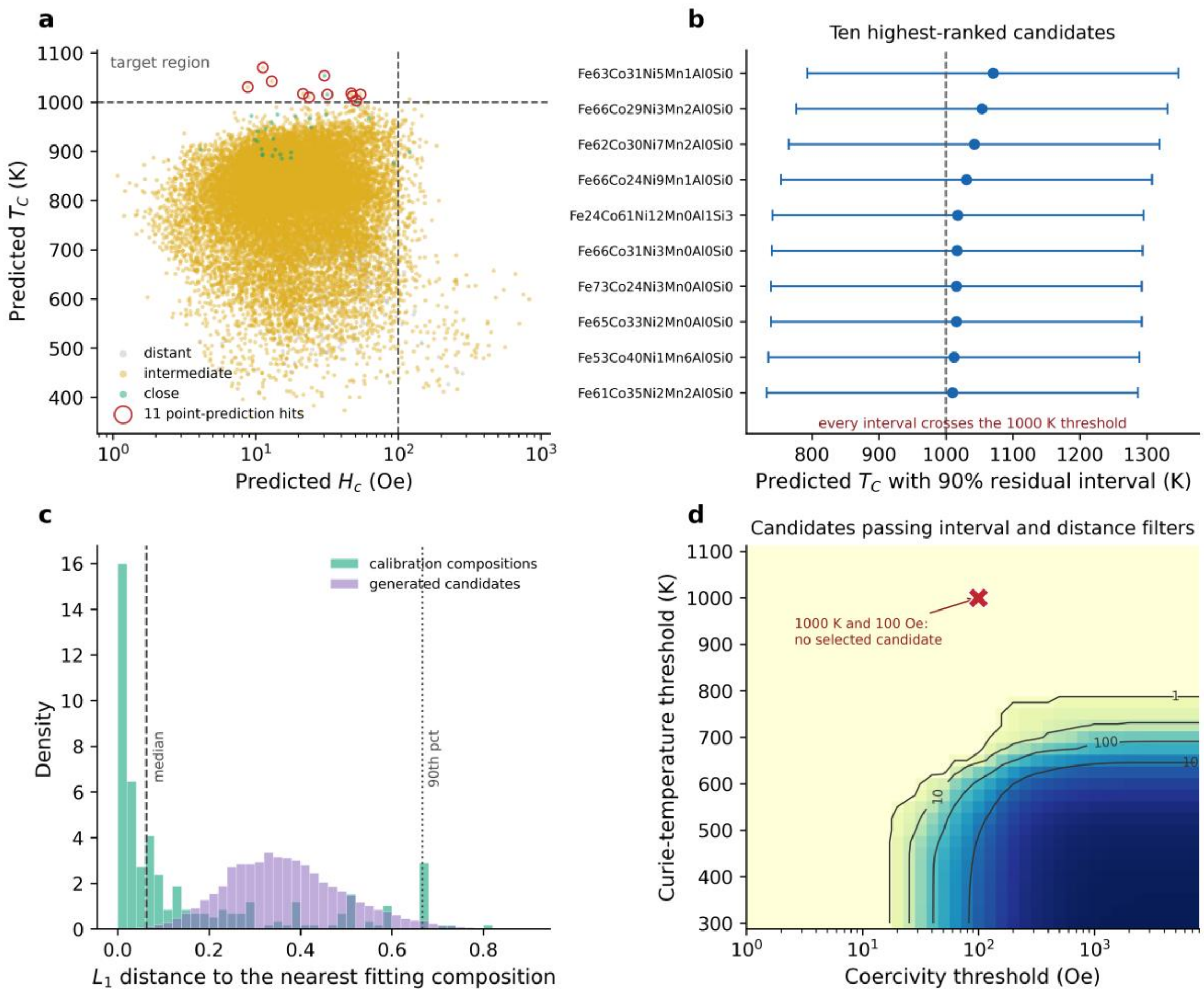

**Figure 5.** Screening of 20,000 Fe–Co–Ni–Mn–Al–Si compositions. (a) Point predictions colored by relative distance to the fitting compositions. Outlined points satisfy the 1,000 K and 100 Oe point criteria. (b) Curie-temperature predictions and nominal 90% residual intervals for the ten highest-ranked candidates. (c) Nearest-composition distances for the calibration set and candidate set; lines indicate the calibration median and 90th percentile. (d) Counts passing both interval criteria and the distance filter over a grid of thresholds.

Candidate counts depend strongly on the selected Curie-temperature and coercivity thresholds. At an upper coercivity threshold of 100 Oe, 899 candidates satisfy both the interval and distance criteria above 500 K, 127 above 600 K, and 3 above 650 K; none satisfies the criteria above 700 K (Figure 5d). Table 3 lists eight of the highest-ranked candidates among the 127 compositions retained at the 600 K and 100 Oe thresholds. The highest-ranked candidate, R1, has a predicted $T_C$ of 940 K with a lower interval bound of 663 K and a predicted $H_C$ of 6.0 Oe with an upper interval bound of 98 Oe. These compositions should therefore be interpreted as prioritized candidates for further evaluation rather than validated soft-magnetic alloys.

**Table 3.** Eight highest-ranked candidates among the 127 passing the interval and distance filters at 600 K and 100 Oe. Parentheses contain the lower Curie-temperature and upper coercivity bounds at the nominal 90% marginal level. Fractions are atomic percent. Distance categories are defined in Section 2.5. Differences smaller than the interval widths are not resolved, and no candidate has been experimentally validated.

| ID | Composition (at.%) | TC (K), lower bound | $H_c$ (Oe), upper bound | L1 distance | Distance class |
|---|---|---|---|---|---|
| R1 | Fe 47.0 Co 24.2 Ni 18.6 Mn 0.0 Al 1.5 Si 8.6 | 940 (663) | 6.0 (98) | 0.191 | Intermediate |
| R2 | Fe 30.7 Co 36.5 Ni 17.4 Mn 0.0 Al 5.1 Si 10.2 | 936 (660) | 3.2 (51) | 0.235 | Intermediate |
| R3 | Fe 44.2 Co 38.1 Ni 7.1 Mn 0.5 Al 5.0 Si 5.0 | 935 (658) | 5.4 (88) | 0.352 | Intermediate |
| R4 | Fe 36.0 Co 44.3 Ni 8.3 Mn 1.1 Al 4.6 Si 5.7 | 926 (649) | 5.6 (91) | 0.394 | Intermediate |
| R5 | Fe 37.7 Co 35.8 Ni 17.6 Mn 0.1 Al 4.6 Si 4.1 | 921 (644) | 5.2 (85) | 0.223 | Intermediate |
| R6 | Fe 29.2 Co 49.7 Ni 7.0 Mn 0.5 Al 6.1 Si 7.5 | 918 (641) | 6.0 (97) | 0.356 | Intermediate |
| R7 | Fe 36.9 Co 35.2 Ni 17.5 Mn 0.9 Al 2.8 Si 6.7 | 918 (641) | 5.8 (95) | 0.222 | Intermediate |
| R8 | Fe 44.2 Co 35.7 Ni 6.3 Mn 3.1 Al 6.5 Si 4.1 | 917 (640) | 6.1 (98) | 0.400 | Intermediate |

The median distance to the fitting compositions is 0.063 for the calibration set and 0.358 for the candidate set (Figure 5c). Most candidate compositions are therefore farther from the fitting data than the typical calibration composition. The distance filter excludes candidates beyond the calibration-set 90th percentile, but it does not guarantee interval coverage or physical feasibility. The selected compositions remain hypotheses for subsequent structural, processing and magnetic evaluation. The Fe-Co-rich character of the highest-ranked candidates is physically plausible but should not be overinterpreted. Slater-Pauling behavior relates magnetic moment to electronic filling and does not establish either a maximum Curie temperature or minimum coercivity at a particular valence electron concentration. Moreover, hardness, ductility, core loss, and electrical resistivity are not included as screening objectives. The present analysis therefore identifies candidates with favorable predicted $T_C$ and $H_C$, rather than application-qualified soft magnets. This distinction is consistent with multi-objective alloy-design approaches that explicitly incorporate additional performance criteria [30,31].

### 3.5 Comparison with reported machine learning studies

Table 4 summarizes representative machine-learning studies of magnetic materials and their evaluation protocols. Direct comparison of reported $R^2$ values is not appropriate because the studies differ in target definitions, data sources, composition ranges, processing variables, and validation strategies. Several studies also include experimental validation of selected candidates, which addresses a different question from the statistical screening performed here.

**Table 4.** Comparison of selected machine-learning studies of magnetic materials, including target properties, data sources, validation protocols, composition-grouping strategies, and reported predictive performance.

| System | Target | Target source | Partition | Exact composition grouping | Reported $R^2$ | Paper |
|---|---|---|---|---|---|---|
| Fe-based nanocrystalline | $H_c$, $M_s$, $\lambda_s$ | experiment | 20-fold CV, random | - | reported graphically | Wang et al. 2020 [43] |
| Fe–Si–Al ternary | $H_c$ and others | experiment | 85:15 random | - | 0.935 ($H_c$) | Milyutin et al. 2024 [32] |
| Fe-based soft magnets | $M_s$, $H_c$ | experiment | random with CV | - | 0.86 / 0.76 | Nachnani et al. 2025 [29] |
| soft magnetic HEAs | phase, $B_0$, M, $T_C$ | DFT | 80:20 random | - | 0.921–0.987 | Dai et al. 2025 [30,31] |
| general ferromagnets | $T_C$ | Compiled data | 80:20 random, 5-fold CV | - | 0.66 / 0.87–0.91 | Abedi Orang et al. 2026 [33] |
| superconductors | log Tc | experiment | random vs leave-one-cluster-out | Chemical clusters | 0.87 vs 0.30 | Meredig et al. 2018 [34] |
| compiled magnetic materials | $T_C$, $\log_{10} H_c$ | Compiled labels | composition-grouped, 20 repeats | Yes (exact compositions) | 0.601 ± 0.080 / 0.589 ± 0.069 | This work |

For example, Milyutin et al. [32] reported $R^2$ = 0.935 for coercivity in a restricted Fe-Si-Al composition space and subsequently tested synthesized candidates. In contrast, the present $R^2$ = 0.589 +/- 0.069 describes logarithmic coercivity across a heterogeneous magnetic-material compilation with entire encoded compositions excluded from training. Similarly, the Curie-temperature results reported by Orang et al. [33] demonstrate the influence of data curation and feature representation on predictive accuracy. These differences emphasize that reported $R^2$ values characterize different prediction tasks and cannot by themselves establish relative model performance. The validation strategy also determines the degree of chemical transfer being tested. The leave-one-cluster-out evaluation of Meredig et al. [34] excludes chemically related groups, whereas the present composition-grouped evaluation excludes exact repeated encoded compositions. Chemically similar compositions may therefore remain in the training data. Publication-level grouping would provide an additional test of transfer across literature sources, but the source identifiers required for such an analysis are not retained in the encoded dataset. Performance under these stronger distribution shifts consequently remains unresolved.

### 3.6 Static Magnetic Benchmarks and Candidate Energetics

The static first-principles volume scans exhibit well-defined sampled energy minima for all four benchmark structures (Figure 6). The lowest sampled energies occur at lattice parameters of 2.830 Å for bcc Fe, 2.850 Å for B2 FeCo, 3.580 Å for $L1_2$ $Fe_3Co$, and 3.540 Å for $L1_2$ $FeCo_3$. The corresponding magnetic moments are 2.168, 2.260, 2.299, and 1.833 $\mu_B$/atom, respectively. These calculations establish energetically preferred regions within the sampled volume ranges together with substantial magnetic polarization in the Fe–Co benchmark structures; they are not interpreted as fitted equilibrium lattice parameters or experimental saturation magnetizations. The pronounced variation in magnetic moment with lattice parameter for $L1_2$ $Fe_3Co$ further indicates strong magnetovolume coupling (Figure 6c). For bcc Fe, the calculation at 2.750 Å converges to a nearly compensated magnetic state from an antiparallel initial spin configuration and is therefore excluded from the ferromagnetic energy branch shown in Figure 6a. Overall the benchmark calculations provide first-principles support for the strong magnetic character of the Fe–Co-rich chemistry emphasized by the machine-learning screening, while also demonstrating the sensitivity of the magnetic state to crystal structure and volume.

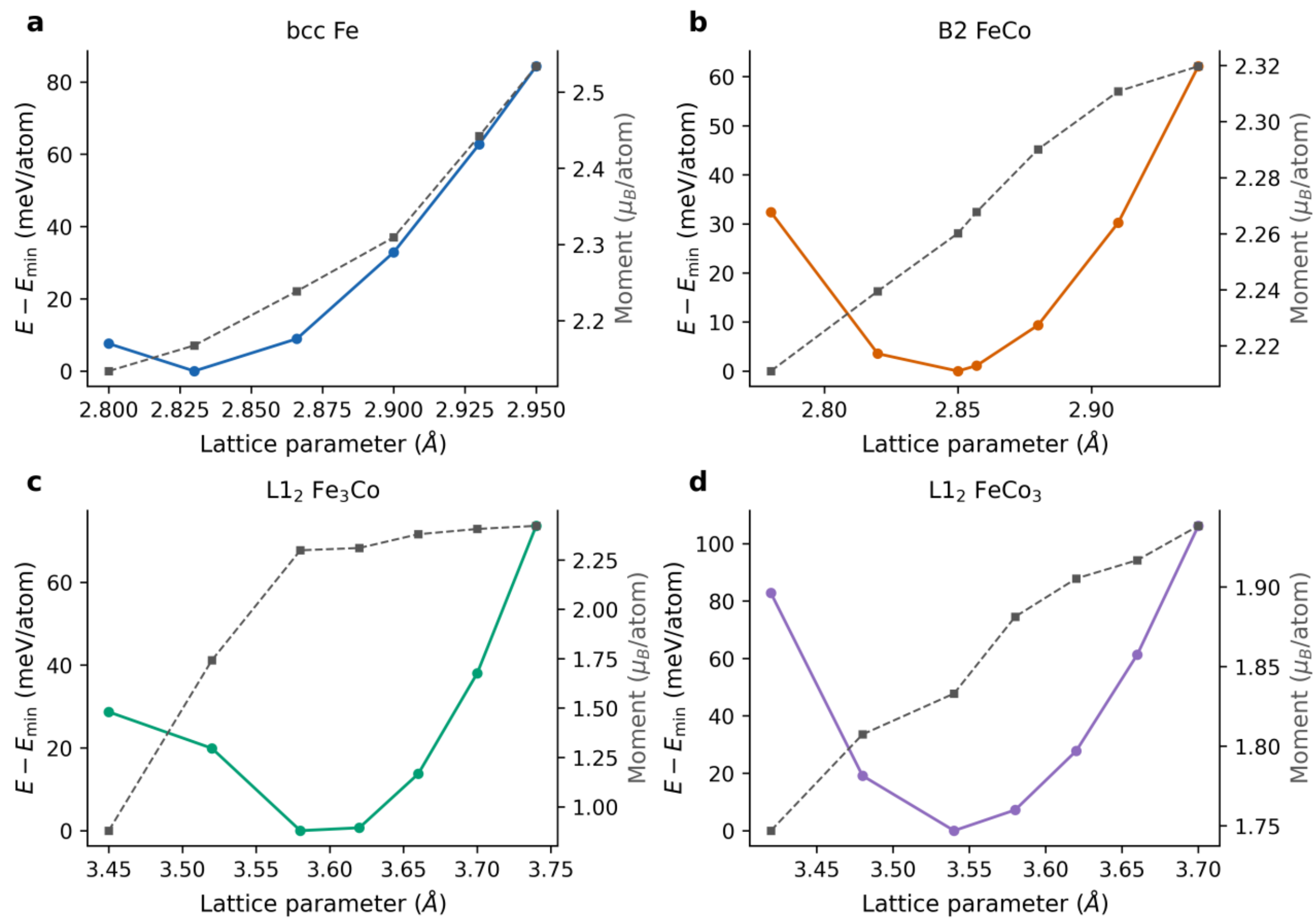

**Figure 6.** Solid colored lines show Eσ→0 relative to the lowest sampled energy for each material, normalized by the actual atom count. Dashed gray lines show the total moment per atom on the right axes. (a) bcc Fe, (b) B2 FeCo, (c) $L1_2$ $Fe_3Co$ and (d) $L1_2$ $FeCo_3$. Lines connect calculated points; no equation-of-state fit is implied. The unlike-spin bcc Fe calculation at 2.750 Å is excluded.

Three relaxed multicomponent BCC/FCC cell pairs were also examined as first-principles representatives of the Fe–Co-rich compositional region identified by the machine-learning screening. Their compositions are similar to those of the highest-ranked screened candidates, although exact matching is limited by the integer site occupancies of the finite supercells. For the three pre-existing candidate-cell pairs, the normalized static FCC-minus-BCC energy differences are +49.35, +61.59, and −5.94 meV/atom for C1, C2, and C3, respectively (Figure 7a). These values cannot be interpreted as phase-stability energies because the paired BCC and FCC cells do not have identical compositions. The composition differences are shown explicitly in Figure 7b. For example, C1 compares $Fe_{25}Co_{18}Ni_5Mn_2AlSi_3$ in the 54-atom BCC cell with $Fe_{49}Co_{36}Ni_{11}Mn_4Al_2Si_6$ in the 108-atom FCC cell. Composition differences are also present for C2 and C3. Because the paired structures differ slightly in composition and a complete competing-phase analysis was not performed, these energy differences do not establish a thermodynamic ground state. The corresponding magnetic moments are 1.867, 1.686, and 1.427 $\mu_B$/atom for the BCC cells and 1.550, 1.442, and 1.252 $\mu_B$/atom for the FCC cells. These results establish magnetic polarization of the calculated states but do not determine coercivity or resolve the compositional differences between the paired structures. A quantitative phase-stability comparison would require matched compositions and consistent structural and magnetic treatment.

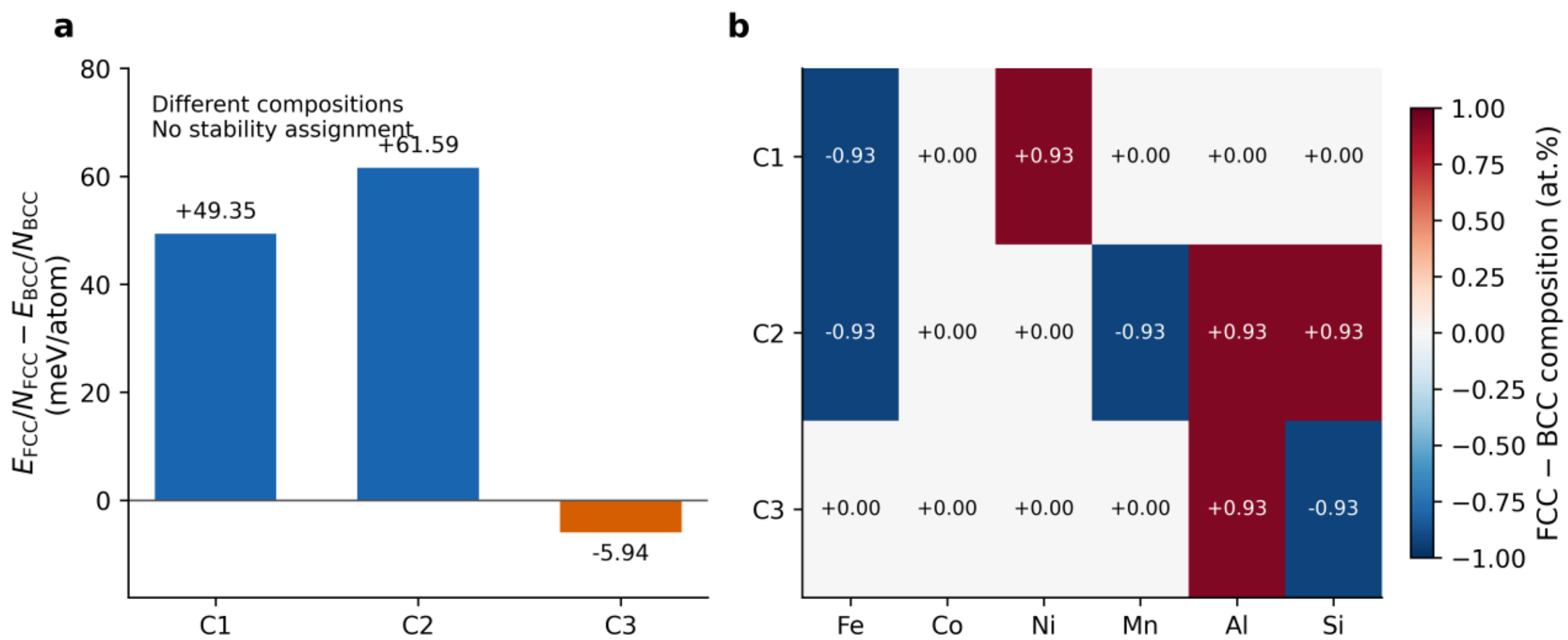

**Figure 7.** (a) Differences between normalized static Eσ→0 values for the relaxed FCC and BCC candidate cells. (b) Actual FCC-minus-BCC elemental fractions in atomic percent. Because the compositions differ, the energy bars are not phase-stability energies. C1–C3 denote the candidate cells, not the generated recipes in Table 3.

### 3.7 Finite-temperature magnetic behavior

Finite-temperature atomistic spin simulations were used to examine the temperature-dependent magnetic behavior of the benchmark systems. Figure 8a shows the normalized magnetization for bcc Fe, B2 FeCo, $L1_2$ $Fe_3Co$, $L1_2$ $FeCo_3$, fcc Co, and fcc Ni. All six systems exhibit a progressive decrease in magnetization with increasing temperature, followed by a transition to a low-magnetization state. Using the first temperature at which the normalized magnetization decreases to 0.1 or below as a consistent transition-scale criterion, the calculated values are approximately 1,110 K for bcc Fe, 1,245 K for B2 FeCo, 1,320 K for $L1_2$ $Fe_3Co$, 1,145 K for $L1_2$ $FeCo_3$, 1,410 K for fcc Co, and 655 K for fcc Ni. These values characterize the Curie-temperature scale of the calculated magnetization curves rather than precise thermodynamic $T_C$ values, because the transitions are identified from magnetization decay rather than susceptibility peaks or finite-size scaling. The Fe-containing systems retain magnetic order to comparatively high temperatures, with B2 FeCo and $L1_2$ $Fe_3Co$ exhibiting higher transition scales than bcc Fe in the present calculations. Together with the static first-principles magnetic moments, these finite-temperature results provide independent physical support for the strong magnetic character of the Fe-Co-rich chemistry emphasized by the machine-learning screening. These benchmark calculations do not constitute candidate-specific validation of the predicted $T_C$ values because the screened six-element compositions were not reproduced exactly; instead, the simulations isolate representative Fe–Co-rich and elemental magnetic responses within the broader candidate composition space. Figure 8b shows the corresponding net-magnetization behavior of fcc Mn. Its net magnetization remains negligible throughout the investigated temperature range, in contrast to the ferromagnetic-like behavior in Figure 8a. Because net magnetization is not an appropriate order parameter for an antiferromagnetic state, no transition temperature is assigned from this curve. Determination of the Néel temperature would require an appropriate antiferromagnetic order parameter or susceptibility-based analysis.

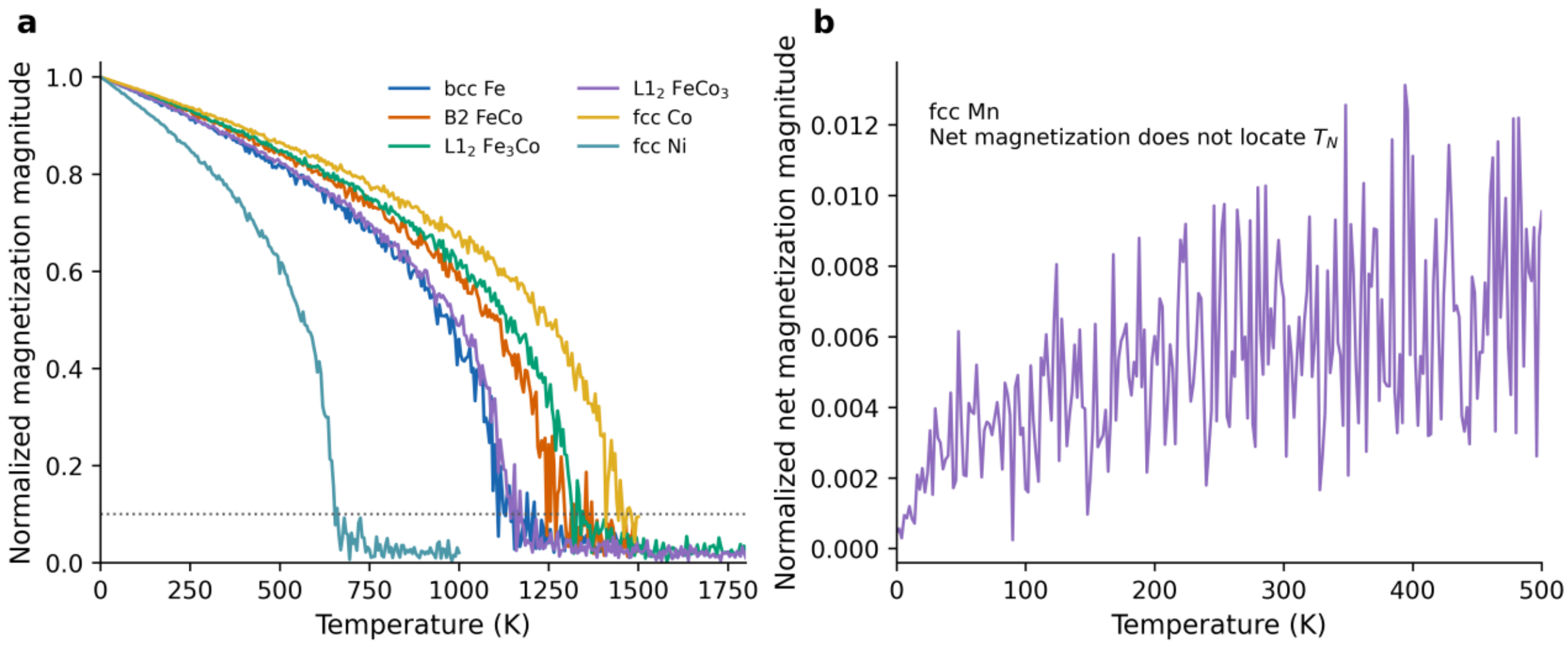

**Figure 8.** Finite-temperature magnetic behavior of the benchmark systems. (a) Normalized magnetization of bcc Fe, B2 FeCo, L1$_2$ $Fe_3Co$, L1$_2$ $FeCo_3$, fcc Co, and fcc Ni as a function of temperature. The dotted line marks the m = 0.1 criterion used to define a consistent transition-temperature scale. (b) Net magnetization of fcc Mn, shown separately because net magnetization does not provide an estimator of the Néel temperature for an antiferromagnetic state.

## 4 Conclusions

A coercivity-aware machine-learning framework was developed for simultaneous screening of Curie temperature and logarithmic coercivity using composition-grouped evaluation with nested model selection. Across twenty grouped partitions, mean test $R^2$ values of 0.601 ± 0.080 and 0.589 ± 0.069 were obtained for Curie temperature and logarithmic coercivity, respectively. Random partitioning produced higher scores but retained substantial composition overlap between training and testing, demonstrating the importance of composition-grouped evaluation when the objective is to screen previously unseen chemistries. In the separate descriptor comparison, elemental fractions gave a mean Curie-temperature $R^2$ of 0.600 ± 0.062, while combining elemental fractions with weighted elemental statistics increased the score to 0.630 ± 0.062. In contrast, elemental fractions alone performed better for logarithmic coercivity than the combined descriptor set, consistent with the strong dependence of coercivity on processing, defects and microstructure.

Residual-based prediction intervals were incorporated to account for predictive uncertainty during candidate selection and achieved mean outer-test coverages of 0.910 and 0.908 for Curie temperature and logarithmic coercivity, respectively. Among 20,000 screened Fe–Co–Ni–Mn–Al–Si compositions, 11 satisfy the point-prediction criteria of $T_C > 1{,}000$ K and $H_C < 100$ Oe, although none satisfies both interval criteria at these thresholds. At $T_C$ > 600 K and $H_C$ < 100 Oe, 127 candidates satisfy both the interval and composition-distance filters, providing a focused set of rare-earth-free chemistries for further investigation.

First-principles and finite-temperature magnetic calculations provide an independent physical assessment of the Fe–Co-rich region emphasized by the machine-learning screening. Static first-principles calculations show substantial magnetic polarization and strong magnetovolume dependence in representative Fe–Co structures. The relaxed multicomponent BCC/FCC cells also retain substantial magnetic moments, although their small compositional differences prevent the calculated energy differences from being interpreted as thermodynamic phase-stability energies. Finite-temperature atomistic spin simulations further show that the Fe–Co benchmark structures retain magnetic order to comparatively high temperatures. These calculations provide independent magnetic benchmarks rather than candidate-specific validation because the screened six-element compositions were not individually simulated. Overall, the combined framework integrates composition-held-out machine learning, uncertainty-aware candidate selection and independent magnetic calculations to prioritize rare-earth-free soft-magnetic chemistries for targeted first-principles investigation and experimental validation.

## Data and code availability

All machine-learning models, alloy design-space generation tools, prediction scripts, figure-generation workflows, DFT input, POSCARS developed in this study are openly available at:
https://github.com/mahata-lab/soft-magnet-ML

## Acknowledgments

The authors acknowledge the use of computational resources at the Massachusetts Green High Performance Computing Center. This research also benefited from high performance computing allocations provided by the National Science Foundation through ACCESS under awards MAT250103 and MAT240094. Additional computational resources were supported by Argonne National Laboratory under the Director's Discretionary allocation for the project GNNMD. Further support was provided through a National Science Foundation Major Research Instrumentation award to Wilkes University under award number 1920129, which contributed essential computational infrastructure for this study.